\documentclass[reprint,aps,prstab,showpacs,amsmath,amssymb,nofootinbib]{revtex4-1}
\usepackage{graphics}
\usepackage{graphicx}
\usepackage{dcolumn}
\usepackage{bm}

\begin{document}

\title{Geometry and space-time extent of pion emission region at FCC energies}

\author{V.A. Okorokov}
\email{VAOkorokov@mephi.ru; Vitaly.Okorokov@cern.ch}
\affiliation{National Research Nuclear University MEPhI
(Moscow Engineering Physics Institute), Kashirskoe shosse 31, 115409 Moscow, Russian Federation}

\date{\today}

\begin{abstract}
The energy dependence is investigated for a wide set of space-time
characteristics derived from Bose\,--\,Einstein correlations of
secondary pion pairs produced in proton-proton and nucleus-nucleus
interactions. Analytic functions
suggested for smooth approximations of the energy dependence of
emission region parameters demonstrate reasonable agreement with all available experimental results for proton-proton collisions while the approximations correspond to most of experimental data for nucleus-nucleus collisions at energies above 5 GeV.
Estimations for a wide set of space-time quantities are obtained for energies
for the Future Circular Collider (FCC) project based on the smooth
approximations. The space particle densities at freeze-out are
derived also from estimations for the volume of the emission region and
for total multiplicity at FCC energies. Estimations for charged particle density and its critical value allow the possibility of lasing behavior for secondary pions in nucleus-nucleus collisions at FCC energy. The mathematical formalism is presented for study of the peak shape of correlation function for general case of central-symmetrical L\'{e}vy\,--\,Feldheim distribution.
\end{abstract}


\maketitle

\section{Introduction}\label{sec.1}
When two energetic particles or nuclei collide, some matter is
created in finite space-time volume. This matter volume, often called ``fireball", emits particles and space-time extent
of the fireball is of fundamental interest for understanding of
both the multiparticle production dynamics and the evolution of
early Universe. One of the collective effects, namely, particle
correlations at low relative momentum represent a unique tool and
sensitive probe of the size and the shape of the fireball at the
last stage of its evolution (colorless particle emission region).
The space-time geometry of particle source can be determined by
using a method of interferometry based on the fundamental relation
between spin and statistics. The production of identical bosons
that are close together in phase space is enhanced by the presence
of quantum statistical effect on Bose\,--\,Einstein correlations
(BEC). The strength and form of the correlation reflects the
space-time structure of the source \cite{Weiner-book-1-2000}. The
most of secondary particles produced in the strong interactions
are pions. Thus in the paper correlations between two identical bosons
called BEC are studied for
secondary charged pions produced in various strong interaction
processes\footnote{In these reactions the BEC are often called HBT
correlations due to analogy with Hanbury-Brown and Twiss effect
\cite{Hanbury-Brown-PhylMag-45-663-1954,*Hanbury-Brown-Nature-177-27-1956,*Hanbury-Brown-Nature-178-1046-1956}
used in radio astronomy to measure the angular sizes of stellar
objects.}.

The international project called Future Circular Collider (FCC) is
mostly aimed at hadron collider with a centre-of-mass energy
$\sqrt{s_{\mbox{\scriptsize{pp}}}}=100$ TeV for $p+p$ collisions
in a new 100 km tunnel of the CERN accelerator complex and
detailed characteristics of various beams for FCC can be found
elsewhere \cite{Schaumann-arXiv-1503.09107}. For heavy-ion
collisions the relation
$\sqrt{s_{\mbox{\scriptsize{NN}}}}=\sqrt{s_{\mbox{\scriptsize{pp}}}
\times (Z_{1}Z_{2}/A_{1}A_{2})}$ gives the energy in
centre-of-mass per nucleon-nucleon collision of
$\sqrt{s_{\mbox{\scriptsize{NN}}}}=39$ TeV for
$\mbox{Pb}+\mbox{Pb}$ ($Z=82$, $A=208$) and 63 TeV for
$p+\mbox{Pb}$ collisions
\cite{Armesto-NPA-931-1163-2014,*Dainese-arXiv-1605.01389,Armesto-arXiv-1601.02963}.
This project provides a unique opportunity to probe quantum
chromodynamics (QCD) in the new energy regime \cite{Chang-SChPMA-59-621001-2016}. The one of the most
distinguishing features of QCD is the mechanism of color
confinement, the physics of which is not fully understood, due, in
part, to its theoretical intractability
\cite{Fodor-JHEP-0302-014-2002}. The confinement mechanism has a
physical scale of the order of the proton radius and is especially
important at low momentum. Therefore study of source geometry in
new energy domain with help of BEC seems important for better
understanding of both the equation of state (EOS) of strongly
interacting matter and general dynamic features of soft processes.
The peak of two-particle Bose\,-–\,Einstein correlation function
(CF) contains the unique experimental information about particle
source at freeze-out. The peak shape carries information, in
particular, about the possible complex highly irregular geometry
of the source
\cite{Okorokov-UJPA-1-196-2013,Okorokov-arXiv-1312.4269}, the
$U_{\mbox{\scriptsize{A}}}$(1) symmetry restoration in high energy
heavy ion reactions \cite{Vance-PRL-81-2205-1998} etc. Therefore
the development of a general formalism for detailed shape analysis
of the peak in BEC CF is relevant for future high-statistics
studies at FCC. It should be stressed that on the one hand the BEC
leads also to Bose\,--\,Einstein condensates responsible for
laser, superfluids and superconductors \cite{Weiner-book-1-2000}.
On the other hand the pion multiplicity at midrapidity
$\left.(dN/d\eta)\right|_{\eta=0}$ is larger than $10^{3}$ in
heavy ion collisions in the TeV-energy domain, in particular, at
FCC the $\left.(dN/d\eta)\right|_{\eta=0} \simeq 3600$ in
$\mbox{Pb}+\mbox{Pb}$ collisions at
$\sqrt{s_{\mbox{\scriptsize{NN}}}}=39$ TeV
\cite{Armesto-NPA-931-1163-2014,*Dainese-arXiv-1605.01389}.
Therefore the number of pions in a unit value of phase-space may
be large enough that these bosons condense into the same quantum
state and a pion laser could be created
\cite{Pratt-PLB-301-159-1993,Pratt-PRC-50-469-1994}. Thus the paper is focused on the
study of azimuthally integrated BEC of secondary charged pions
produced in strong interactions, especially, on the space-time
extent of pion emission region and the possible novel features of
multiparticle production mechanism (pion laser) at FCC energies.
Also the general formalism is suggested for study of shape of
correlation peak in detail.

The paper is organized as follows. In Sec.~\ref{sec.2}, definitions of two-particle CF and BEC parameters are described.
The Sec.~\ref{sec.3} devotes discussion of energy dependence of pion source extent in $p+p$ and $\mbox{A}+\mbox{A}$ collisions, predictions for wide set of space-time characteristics for pion source in various collisions at FCC energies. The possibility for pion laser in strong interaction processes at FCC energies is considered in Sec.~\ref{sec.4}. In Sec.~\ref{sec.5}, the generalized parametrization of 3D CF is introduced with help of expansion of central-symmetrical L\'{e}vy\,--\,Feldheim distribution. Some final remarks are presented in Sec.~\ref{sec.6}.

\section{Method and variables}\label{sec.2}
The BEC effect is observed as an enhancement in the two-particle
CF at low values of some difference constructed from 4-momenta
$p_{i}$, $i=1,2$ or its components of the entering particles,
$C_{2}(p_{1},p_{2}) = \rho(p_{1},p_{2}) /
\rho_{\scriptsize{\mbox{ref}}}(p_{1},p_{2})$, where the $\rho$ is the two-particle density function,
$\rho_{\scriptsize{\mbox{ref}}}$ is a reference two-particle
density function that by construction is expected to include no
BEC. Recent study \cite{Okorokov-arXiv-1605.02927} shows that BEC 1D experimental data samples are
not enough for study of energy dependence of source parameters in proton-nucleus and nucleus-nucleus collisions. Therefore the present paper is focused on the 3D analysis of BEC in strong interaction processes.

In general phenomenological parametrization of CF with taking into account different forms of
corrections on Coulomb final state interaction (FSI) can be
written as follows \cite{Okorokov-arXiv-1312.4269}:
\begin{subequations}
\begin{equation}
C_{2,(m)}^{\mbox{\scriptsize{ph}}}(q,K)=\zeta
P_{\mbox{\footnotesize{coul}}}^{(m)}(q)
\bigl[\zeta^{-1}+\mathbf{K}_{2}^{\mbox{\scriptsize{ph}}}({\bf
A})\bigr],\label{eq:2.1a}
\end{equation}
\vspace*{-0.5cm}
\begin{equation}
\mathbf{K}_{2}^{\mbox{\scriptsize{ph}}}({\bf
A})=C_{2,(m)}^{\mbox{\scriptsize{ph}}}(q,K)-1,
\end{equation}\label{eq:2.1}
\end{subequations}
\hspace*{-0.1cm}where $\mathbf{K}_{2}$ is the cumulant correlation function (cCF), $\zeta = \lambda$ at $m=1,2$ and $\zeta=1$ at $m=3$ while
$m=1$ corresponds to the standard Coulomb correction, $m=2$
-- the dilution procedure and $m=3$ -- the Bowler\,--\,Sinyukov
correction, $q \equiv (q^{0},\vec{q})=p_{\,1}-p_{\,2}$ is the
relative 4-momentum, $K \equiv
(K^{0},\vec{K})=(p_{\,1}+p_{\,2})/2$ -- the average 4-momentum of
particles in pair (pair 4-momentum), ${\bf A} \equiv \vec{q}\,{\bf R}^{2}\vec{q}^{\,T}$ and ${\bf
R}^{2}$ are the matrices $3 \times 3$, $\vec{q}^{\,T}$ --
transposed vector $\vec{q}$, $\forall~ i,j=1-3:
R^{\,2}_{ij}=R^{\,2}_{ji}, R^{\,2}_{ii} \equiv R^{\,2}_{i}$, where
$R_{i}=R_{i}(K)$ are parameters characterized the linear scales of the region of
homogeneity \cite{Sinyukov-NATOSeries-346-309-1995,*Sinyukov-PLB-356-525-1995}; the
products are taken on space components of vectors,
$\lambda(K)=\mathbf{K}_{2}(0,K), 0 \leq \lambda \leq 1$ is the
parameter which characterizes the strength of correlations called also chaoticity.
Different types of Coulomb correction for two-pion correlations
are compared in \cite{Okorokov-arXiv-1312.4269}. The space
component of pair 4-momentum ($\vec{K}$) is decomposed on
longitudinal
$k_{\parallel}=(p_{\,\parallel,1}+p_{\,\parallel,2})/2$ and
transverse
$\vec{k}_{\perp}=(\vec{p}_{\perp,1}+\vec{p}_{\perp,2})/2$ parts of
pair momentum. In the paper the decomposition of
Pratt\,--\,Bertsch
\cite{Pratt-PRD-33-1314-1986,*Bertsch-PRC-37-1896-1988} is used
for $\vec{q}$ as well as the longitudinal co-moving system (LCMS)
frame. The parametrization of the $\mathbf{K}_{2}^{\mbox{\scriptsize{ph}}}({\bf A})$
depends on type of distribution which
was chosen for emission region \cite{Okorokov-arXiv-1312.4269}. For
instance, the lowest order cCF can be written as
\begin{equation}
\mathbf{K}_{2,\mbox{\scriptsize{G}}}^{\mbox{\scriptsize{ph}},0}({\bf
A})=
\exp\biggl(-\sum\limits_{i,j=1}^{3}q_{i}R_{ij}^{\,2}q_{j}\biggr).
\label{eq:2.2}
\end{equation}
for specific case of Gaussian distribution which is one of the
most used in BEC study. As known the study of BEC allows the
estimation of space-time extent for region of homogeneity which is
only some part of whole source. Therefore the BEC parameters
$\forall~ i=1-3: R_{i}(K)$ are smaller \emph{a priori} than
corresponding scales of whole emission region and consequently the
experimental BEC dimensions $R_{i}(K)$ can be considered as low
boundary for corresponding true linear scales of source
\begin{equation}
\forall~ i=1-3: R_{i}(K)=\inf R_{i}^{\mbox{\scriptsize{tr}}}(K). \label{eq:2.dop}
\end{equation}
For this reason the BEC parameters $\forall~ i=1-3: R_{i}(K)$ are called BEC radii and it is assumed
that correlation analysis for pairs of identical particles with
low $\langle \vec{k}_{\perp}\rangle$ provides the $R_{i}$ which
are adequate experimental estimations for space-time extent of
whole emission region within the simplest approach at least. It
should be noted that azimuthally integrated BEC analysis allows
rougher estimations for space-time scales of whole source with
increase of collision energy because more intensive collective
expansions reduce the sizes of the region of homogeneity more
significantly at higher energies. Thus in the present paper the
$R_{i}$, $i=1-3$ are considered as source BEC radii with taking
into account the relation (\ref{eq:2.dop}) and influence of
collective flows on the quality of this approximation.

In the 3D case and the Pratt\,--\,Bertsch coordinate system the
space-time extents of the region of homogeneity or, with taking
into account the discussion above, whole source is described by
the following dimensions: $R_{\mbox{\scriptsize{l}}}$ is the
source size along the beam axis, $R_{\mbox{\scriptsize{o}}}$ --
extent along the $\vec{k}_{\perp}$ and $R_{\mbox{\scriptsize{s}}}$
is the source size along the axis perpendicular to those two. Then
one can define the geometric mean BEC radius
\begin{equation}
R_{\mbox{\scriptsize{m}}}^{3} =
\prod_{i=\mbox{\scriptsize{s,o,l}}}R_{i} \label{eq:2.3}
\end{equation}
as well as the difference
\begin{equation}
\delta \equiv
R_{\mbox{\scriptsize{o}}}^{2}-R_{\mbox{\scriptsize{s}}}^{2}
\label{eq:2.4}
\end{equation}
which is an important observable especially for some specific cases
of 1D hydrodynamics (static, nonflowing source) due to its relation with particle emission
duration $\delta \approx \beta_{\perp}^{2}(\Delta\tau)^{2}$
\cite{Wiedemann-PR-319-145-1999,Okorokov-UchPosob-2009}, where
$\beta_{\perp}=k_{\perp}/m_{\perp}$ is the transverse
velocity of pair of particles with mass $m$,
$m_{\perp}^{2}=k_{\perp}^{2}+m^{2}$. Here the scaled geometric
mean BEC radius is defined as follows
$R_{\mbox{\scriptsize{m}}}^{n}=R_{\mbox{\scriptsize{m}}}/R_{\mbox{\scriptsize{A}}}$
in accordance with approach suggested in
\cite{Okorokov-arXiv-1312.4269,Okorokov-AHEP-2015-790646-2015},
where $\langle
R_{\mbox{\scriptsize{A}}}\rangle=(R_{\mbox{\scriptsize{A}}_{1}}+R_{\mbox{\scriptsize{A}}_{2}})/2$
is the mean radius for beam nuclei,
$R_{\mbox{\scriptsize{A}}}=r_{0}A^{1/3}$ is radius of
spherically-symmetric nucleus, $r_{0}=(1.25 \pm 0.05)$ fm
\cite{Valentin-book-1982,*Mukhin-book-1983}. The volume of source
can be written as follows:
\begin{equation}
V=\left\{
\begin{array}{ll}
\vspace*{0.2cm}(2\pi)^{3/2}R_{\mbox{\scriptsize{s}}}^{2}R_{\mbox{\scriptsize{l}}},
& ~~~~~\mbox{(a)} \\
4\pi R_{\mbox{\scriptsize{m}}}^{3}/3, & ~~~~~\mbox{(b)}
\end{array}
\right.\label{eq:2.5}
\end{equation}
where the first case is the standard relation for BEC while the
second case corresponds to the simplest approach of spherically
symmetric source and it can be useful for future study of pion
laser. Thus in the paper the following set of main BEC observables
$\mathcal{G}_{1} \equiv
\{\mathcal{G}_{1}^{i}\}_{i=1}^{4}=\{\lambda,
R_{\mbox{\scriptsize{s}}}, R_{\mbox{\scriptsize{o}}},
R_{\mbox{\scriptsize{l}}}\}$ is under consideration as well as the
set of important additional observables which can be calculated
with help of BEC radii $\mathcal{G}_{2} \equiv
\{\mathcal{G}_{2}^{j}\}_{j=1}^{4}=\{R_{\mbox{\scriptsize{o}}}/
R_{\mbox{\scriptsize{s}}}, R_{\mbox{\scriptsize{m}}}, \delta,
V\}$. The set of parameters $\mathcal{G}_{1}$ characterizes the
chaoticity of source and its 4-dimensional geometry at freeze-out
stage completely.

\section{Space--time extent of pion source}\label{sec.3}
In this study experimental BEC data sets are from
\cite{Okorokov-arXiv-1504.08336} for $p+p$ and from
\cite{Okorokov-AHEP-2015-790646-2015} for $\mbox{A}+\mbox{A}$
collisions.

\begin{figure}[b]
\resizebox{0.45\textwidth}{!}{%
\includegraphics[width=7.0cm,height=8.0cm]{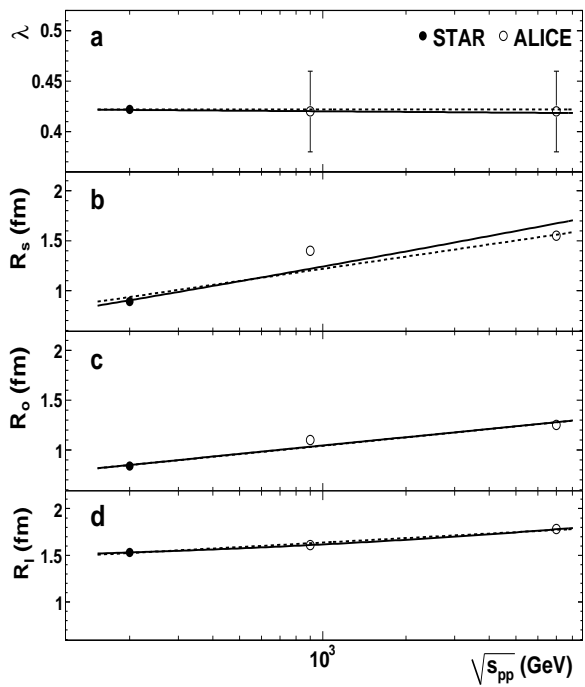}
}\caption{\label{fig:3.1} Energy dependence of the main BEC
parameters $\mathcal{G}_{1}^{i}$, $i=1-4$ obtained for 3D Gaussian
analyses for $p+p$ collisions at midrapidity and low $\langle
k_{T}\rangle \simeq 0.2$ GeV/$c$. Errors for experimental points
are statistical only. For strength of correlations (a) fit by
specific case of (\ref{eq:3.1}) at $a_{3}=1.0$ is shown by solid
line, the dashed line corresponds to the fit by constant. For BEC
radii (b--d) approximations by function (\ref{eq:3.1}) are shown
by solid lines for case (i.1), dashed lines present the results
for linear logarithmic function in the case (i.2).}
\end{figure}

Dependencies of BEC parameters
$\mathcal{G}_{1}^{i}(\sqrt{\smash[b]{s_{\footnotesize{NN}}}})$,
$i=1-4$ for $p+p$ high energy collisions are shown in
Figs.~\ref{fig:3.1}a--d respectively. As seen for energy range from Relativistic Heavy Ion Collider (RHIC) to the Large Hadron Collider (LHC) the experimental
$\lambda(\sqrt{\smash[b]{s_{\footnotesize{NN}}}})$ is close to the
constant (Fig.~\ref{fig:3.1}a) while the some decrease is observed for experimental $\lambda(\sqrt{\smash[b]{s_{\footnotesize{NN}}}})$ deduced from 1D two-pion BEC analyses \cite{Alexander-JPCS-675-022001-2016,*Alexander-arXiv-1610.05298}. The BEC radii increase with
collision energy (Fig.~\ref{fig:3.1}b--d) more significantly in
transverse plane with respect of the beam direction than that for
longitudinal direction. Taking into account the view of
experimental
$\mathcal{G}_{1}^{i}(\sqrt{\smash[b]{s_{\footnotesize{NN}}}})$,
$i=1-4$ in $p+p$ as well as the detailed study of energy
dependence of azimuthally integrated main BEC parameters
$\mathcal{G}_{1}^{i}$, $i=1-4$ for charged pions in
nucleus-nucleus interactions \cite{Okorokov-AHEP-2015-790646-2015}
the following function is used
\begin{equation}
f(\varepsilon) = a_{1}\left[1 +
a_{2}(\ln\varepsilon)^{a_{3}}\right] \label{eq:3.1}
\end{equation}
for smooth approximation of experimental dependencies
$\mathcal{G}_{1}^{i}(\sqrt{\smash[b]{s_{\footnotesize{pp}}}})$,
$i=1-4$ in $p+p$ interactions. Here $\varepsilon \equiv
s_{\footnotesize{pp}}/s_{0}$ for $p+p$ or $\varepsilon \equiv
s_{\footnotesize{NN}}/s_{0}$ for $\mbox{A}+\mbox{A}$ reactions,
where $s_{0}=1$ GeV$^{2}$. The very limited ensemble of
experimental points from 3D Gaussian analyses of $p+p$ does not
allow the fit by (\ref{eq:3.1}) with all parameters
$\forall\,i=1-3: a_{i}$ to be free. In general the various quantities from the set $\mathcal{G}_{1}$ can show
the different behavior as function of $\ln\varepsilon$
\cite{Okorokov-AHEP-2015-790646-2015}. Thus following two views of
(\ref{eq:3.1}) are used for approximation of experimental points:
(i.1) function (\ref{eq:3.1}) at fixed value of $a_{3}$ which is
defined by method of sequential approximations, (i.2) the specific
case of (\ref{eq:3.1}) at $a_{3}=1.0$. Only statistical
uncertainties are available for strength of correlations $\lambda$
while for each BEC radii $\{\mathcal{G}_{1}^{i}\}_{i=2}^{4}$ fits
are made for both the statistical and total errors, where total
errors of experimental points include available clear indicated
systematic errors added in quadrature to statistical ones. The
numerical values of fit parameters are presented in
Table~\ref{tab:3.1}, where the second line for chaoticity
parameter $\lambda$ corresponds to the simplest fit by constant
and for each BEC radii -- to the approximation by specific case of
(\ref{eq:3.1}). Approximation curves are shown in
Fig.~\ref{fig:3.1}a by solid line for specific case of
(\ref{eq:3.1}) and by dashed line for fit by constant.
Figs.~\ref{fig:3.1}b--d show the fit results for BEC radii by solid
lines for approach (i.1) and by dashed lines for specific case
(i.2) with taking into account the statistical errors of
experimental points.

\begin{table*}
\caption{\label{tab:3.1}Values of fit parameters for approximation of $p+p$ data}
\begin{ruledtabular}
\begin{tabular}{lcccccccc}
\multicolumn{1}{l}{BEC} & \multicolumn{4}{c}{Fit with
statistical errors} &
\multicolumn{4}{c}{Fit with total errors} \\
\cline{2-9}
parameter & $a_{1}$ & $a_{2}$ & $a_{3}$ & $\chi^{2}/\mbox{n.d.f.}$ & $a_{1}$ & $a_{2}$ & $a_{3}$ & $\chi^{2}/\mbox{n.d.f.}$ \rule{0pt}{10pt}\\
\hline
$\lambda$                   & $0.43 \pm 0.06$     & $-0.001 \pm 0.012$ & $1.0$  & $7.0 \times 10^{-4}/1$ & --                  & --              & --     & -- \\
                            & $0.422 \pm 0.004$   & --                 & --     & $4.9 \times 10^{-4}/1$ & --                  & --              & --     & --        \\
$R_{\mbox{\scriptsize{s}}}$ & $(6.8 \pm 1.0)\times10^{-3}$ & $7.57 \pm 0.24$    & $1.21$ & $47.7/1$   & $(8.0 \pm 1.2)\times10^{-3}$ & $7.3 \pm 1.0$   & $1.16$ & $0.57/1$ \\
                            & $(2.5 \pm 0.7)\times10^{-3}$ & $35 \pm 7$         & $1.0$  & $62.6/1$   & $(1.6 \pm 1.0)\times10^{-3}$ & $53 \pm 17$     & $1.0$  & $0.72/1$  \\
$R_{\mbox{\scriptsize{o}}}$ & $0.19 \pm 0.08$     & $0.36 \pm 0.07$    & $0.97$ & $3.64/1$   & $0.028 \pm 0.019$   & $4.1 \pm 2.9$   & $0.84$ & $0.11/1$ \\
                            & $0.21 \pm 0.08$     & $0.29 \pm 0.14$    & $1.0$  & $3.71/1$   & $0.19 \pm 0.05$     & $0.33 \pm 0.10$ & $1.0$  & $0.12/1$  \\
$R_{\mbox{\scriptsize{l}}}$ & $1.45 \pm 0.03$     & $(9.2 \pm 1.4) \times 10^{-5}$ & $2.72$  & $9.4 \times 10^{-9}/1$ & $1.40 \pm 0.12$ & $(5.7 \pm 2.3) \times 10^{-4}$ & $2.14$ & $5.7 \times 10^{-4}/1$ \\
                            & $1.16 \pm 0.07$     & $0.030 \pm 0.006$  & $1.0$  & $0.35/1$   & $1.2 \pm 0.6  $     & $0.03 \pm 0.06$ & $1.0$  & $4.9\times10^{-4}/1$ \\
\end{tabular}
\end{ruledtabular}
\end{table*}

In difference with $\mbox{A}+\mbox{A}$ collisions
\cite{Okorokov-AHEP-2015-790646-2015} function (\ref{eq:3.1}) for
both choices of $a_{3}$ agrees with experimental $p+p$ data
quantitatively and provides reasonable fit qualities even with
statistical errors for all BEC parameters from the set
$\mathcal{G}_{1}$ with exception of $R_{\mbox{\scriptsize{s}}}$.
In the last case one can only conclude that fit curve is similar to
the general trend of experimental points (Fig.~\ref{fig:3.1}b)
due to poor fit quality. Account for total errors allows statistically acceptable fit qualities for all main BEC parameters
in both approaches (i.1) and (i.2). Furthermore the $a_{2}$ is
equal to zero within errors for longitudinal BEC radius for (i.2)
and consequently the
$R_{\mbox{\scriptsize{l}}}(\sqrt{\smash[b]{s_{\footnotesize{pp}}}})$
can be described by constant with $a_{1}=(1.57 \pm 0.13)$ fm,
$\chi^{2}/\mbox{n.d.f.}=0.42/2$ in the case of the accounting for total errors. For energy range from RHIC to the LHC fit
curves for approaches (i.1) and (i.2) are close to each other for
$\mathcal{G}_{1}^{i}$, $i=2-4$ especially for radii
$R_{\mbox{\scriptsize{o}}}$ (Fig.~\ref{fig:3.1}c) and
$R_{\mbox{\scriptsize{l}}}$ (Fig.~\ref{fig:3.1}d). Nevertheless
the fit within approach (i.1) for $R_{\mbox{\scriptsize{o}}}$ only
confirms the $R_{\mbox{\scriptsize{o}}} \propto \ln\varepsilon$
for statistical errors but other BEC radii show the faster
increasing with $\sqrt{\smash[b]{s_{\footnotesize{pp}}}}$
especially the $R_{\mbox{\scriptsize{l}}}$ (Table~\ref{tab:3.1}).
The growth of the $R_{\mbox{\scriptsize{s}}}$ approaches to the
linear behavior in $\ln\varepsilon$ for accounting for total
errors but longitudinal radius $R_{\mbox{\scriptsize{l}}}$
preserves much faster growth with energy increase in this case
too. As consequence the method of sequential approximations for
$a_{3}$ leads to significant improvement of the fit quality with
respect to the quantity for approach (i.2) for
$R_{\mbox{\scriptsize{s}}}$ at statistical errors and especially
for $R_{\mbox{\scriptsize{s}}}$ at all considered types of errors.
It should be noted the difference between smooth curves obtained
within the approaches (i.1) and (i.2) for BEC radii can be much
more noticeable for higher energy FCC than that in
Fig.~\ref{fig:3.1} which can have a relevant effect on the estimated BEC parameters at
FCC.

\begin{figure}[b]
\resizebox{0.45\textwidth}{!}{%
\includegraphics[width=7.0cm,height=8.0cm]{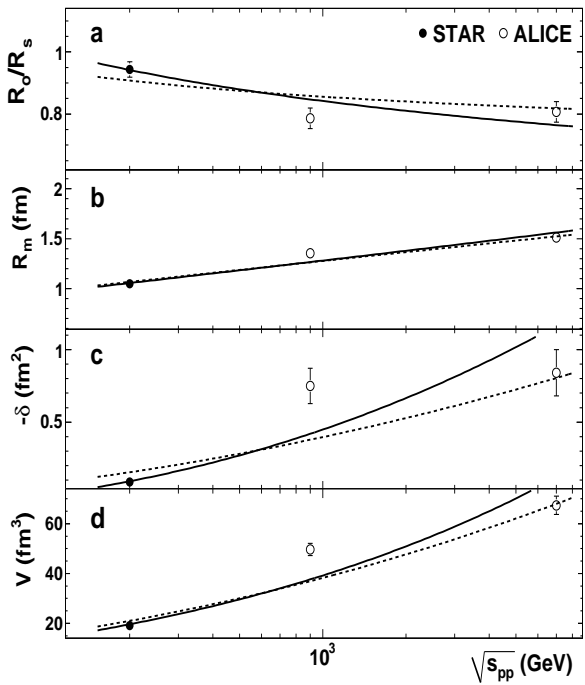}
}\caption{\label{fig:3.2} Energy dependence of the additional BEC
parameters $\mathcal{G}_{2}^{j}$, $j=1-4$ for $p+p$ collisions at
midrapidity and low $\langle k_{T}\rangle \simeq 0.2$ GeV/$c$.
Errors for experimental points are statistical only. Smooth curves
are calculated with help of fit results for BEC radii, solid lines
are from fits of $R_{i}$, $i=\mbox{s}, \mbox{o}, \mbox{l}$ by
function (\ref{eq:3.1}) in the case (i.1) and dashed lines
corresponds to the fits by specific case $R_{i} \propto
\ln\varepsilon$, $i=\mbox{s}, \mbox{o}, \mbox{l}$.}
\end{figure}

Dependencies of additional BEC parameters
$\mathcal{G}_{2}^{j}(\sqrt{\smash[b]{s_{\footnotesize{NN}}}})$,
$j=1-4$ for $p+p$ interactions are shown in
Figs.~\ref{fig:3.2}a--d respectively. Notations of smooth curves
correspond to the Fig.~\ref{fig:3.1} namely the solid curves show
the results calculated with help of fits of BEC radii within
approach (i.1) and dashed lines are for special fits (i.2). As
seen curves of both types agree with experimental points
reasonably for $R_{\mbox{\scriptsize{o}}}/
R_{\mbox{\scriptsize{s}}}$ (Fig.~\ref{fig:3.2}a) and
$R_{\mbox{\scriptsize{m}}}$ (Fig.~\ref{fig:3.2}b) in total
experimentally available energy range. Otherwise the approach
(i.1) leads to significant overpredictions for $-\delta$
(Fig.~\ref{fig:3.2}c) and $V$ (Fig.~\ref{fig:3.2}d) at the LHC
energy $\sqrt{\smash[b]{s_{\footnotesize{NN}}}}=7$ TeV while the
curves for special case (i.2) agree with experimental points at
this energy.

Thus there is significant uncertainty in functional behavior of
dependence of experimental BEC parameters on collision energy due
to very limited ensemble of 3D experimental data for $p+p$ and
future experimental results are important crucially for more
definitive conclusion with regard of behavior of dependencies
$\mathcal{G}_{1,2}^{i}(\sqrt{\smash[b]{s_{\footnotesize{NN}}}})$,
$i=1-4$.

The pion emission duration $\Delta\tau$ for $p+p$ collisions can
be estimated with taking into account the results for $|\delta|$
and kinematic regime for pion pairs under study. The $\langle
\beta_{\perp}\rangle \approx 0.82$ for pion pairs with $\langle
k_{\perp}\rangle \simeq 0.2$ GeV/$c$ as well as for nuclear
collisions \cite{Okorokov-AHEP-2015-790646-2015}. Then pion
emission duration increases from $\Delta\tau=(0.36 \pm 0.08)$
fm/$c$ at RHIC energy
$\sqrt{\smash[b]{s_{\footnotesize{pp}}}}=0.2$ TeV up to the
$\Delta\tau=(1.25 \pm 0.12)$ fm/$c$ at the LHC energy
$\sqrt{\smash[b]{s_{\footnotesize{pp}}}}=7$ TeV which is highest for available experimental BEC results. Thus
the pion emission durations in $p+p$ collisions are smaller significantly
than that for nuclear interactions
\cite{Okorokov-AHEP-2015-790646-2015} in the energy range from RHIC top up to the LHC.

\begin{figure}[h]
\resizebox{0.45\textwidth}{!}{%
\includegraphics[width=7.0cm,height=6.5cm]{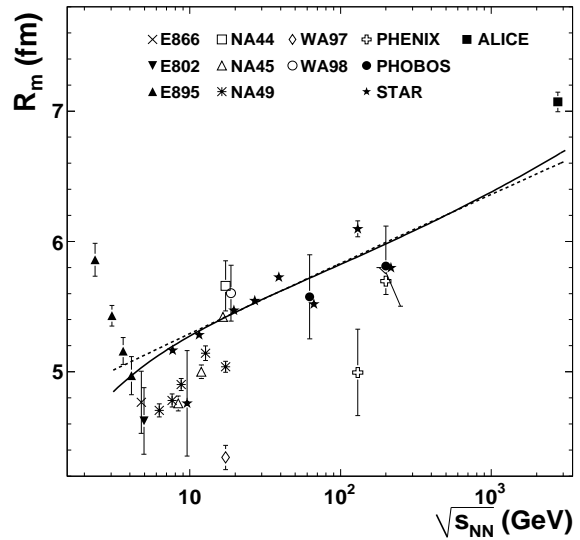}
}\caption{\label{fig:3.3} Energy dependence of
$R_{\mbox{\scriptsize{m}}}$ for secondary charged pions in central
heavy ion collisions $\mbox{Au+Au, Au+Pb, Pb+Pb}$ in midrapidity
region and at $\langle k_{\perp}\rangle \simeq 0.2$ GeV/$c$. Error
bars are only statistical (for NA44, total uncertainties).
Smooth curves are derived from (\ref{eq:2.3}) and the fit
results for BEC radii \cite{Okorokov-AHEP-2015-790646-2015}
without the point of the WA97 experiment
\cite{Antinori-JPGNPP-27-2325-2001}. The solid line corresponds to
the fits of BEC radii by function (\ref{eq:3.1}) and dashed line
-- to the fits by specific case $\forall\,i=\mbox{s}, \mbox{o},
\mbox{l}: R_{i} \propto \ln\varepsilon$.}
\end{figure}

\begin{figure}[h]
\resizebox{0.44\textwidth}{!}{%
\includegraphics[width=7.3cm,height=6.6cm]{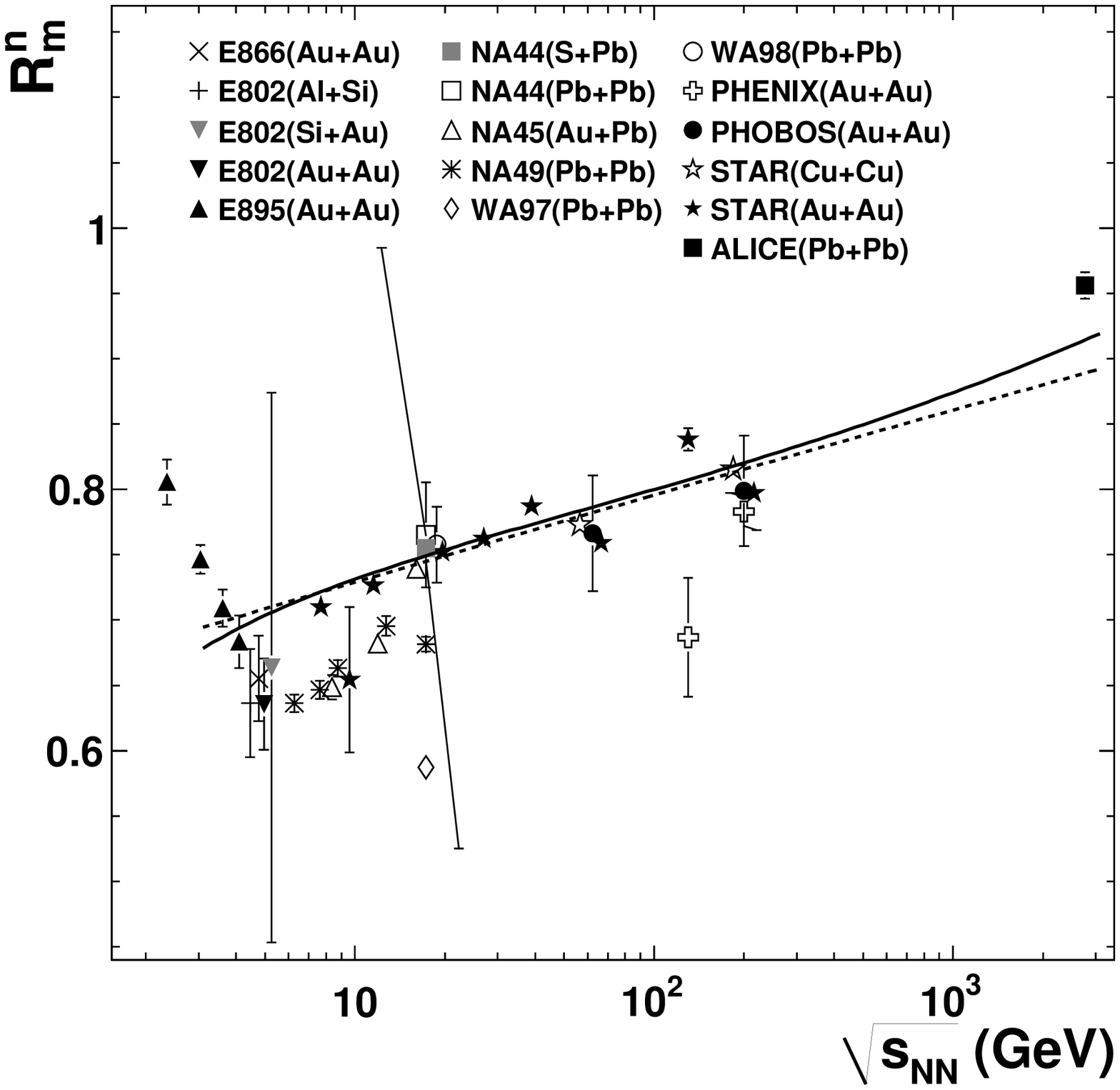}
}\caption{\label{fig:3.4} Energy dependence of scaled
$R_{\mbox{\scriptsize{m}}}^{n}$ for secondary charged pions in
various nucleus-nucleus collisions at $\langle k_{\perp}\rangle
\simeq 0.2$ GeV/$c$. Experimental results are shown for central
collisions (for minimum bias event in the case of E802 for
$\mbox{Al+Si}$), for pairs of $\pi^{-}$ mesons (in the cases of
ALICE and STAR for both the $\mbox{Cu+Cu}$ and $\mbox{Au+Au}$ at
$\sqrt{\smash[b]{s_{\footnotesize{NN}}}}=7.7-62.4$ and 200 GeV,
for $\pi^{\pm}\pi^{\pm}$ pairs, E802 for $\mbox{Al+Si}$, NA44 for
$\mbox{S+Pb}$, for pairs of $\pi^{+}$ mesons) and for standard
Coulomb correction $P_{\mbox{\scriptsize{C}}}^{(1)}(q)$ (in the
cases of ALICE, NA44, NA45, PHOBOS, STAR for both the
$\mbox{Cu+Cu}$ and $\mbox{Au+Au}$ at
$\sqrt{\smash[b]{s_{\footnotesize{NN}}}}=7.7, 11.5-62.4$ and 200
GeV, for correction $P_{\mbox{\scriptsize{C}}}^{(3)}$).
Statistical errors are shown (for NA44, total uncertainties).
Smooth curves are derived from fit results for scaled BEC radii
\cite{Okorokov-AHEP-2015-790646-2015}. The solid line corresponds
to the fits of scaled BEC radii by function (\ref{eq:3.1}) and
dashed line -- to the fits by specific case $\forall\,i=\mbox{s},
\mbox{o}, \mbox{l}: R_{i}^{n} \propto \ln\varepsilon$.}
\end{figure}

The energy dependence for BEC parameters from the sets
$\mathcal{G}_{1,2}^{i}(\sqrt{\smash[b]{s_{\footnotesize{NN}}}})$,
$i=1-4$ was studied in detail in
\cite{Okorokov-AHEP-2015-790646-2015} for nuclear collisions with
exception of the $R_{\mbox{\scriptsize{m}}}$ and corresponding
scaled quantity $R_{\mbox{\scriptsize{m}}}^{n}$ defined in the
present paper. Energy dependence of these additional BEC
parameters are obtained with experimental database for nuclear
collisions from \cite{Okorokov-AHEP-2015-790646-2015}.
Figs.~\ref{fig:3.3}, \ref{fig:3.4} show the experimental
$R_{\mbox{\scriptsize{m}}}(\sqrt{\smash[b]{s_{\footnotesize{NN}}}})$
and
$R_{\mbox{\scriptsize{m}}}^{n}(\sqrt{\smash[b]{s_{\footnotesize{NN}}}})$
respectively as well as smooth curves calculated with fit results
for BEC radii at $\sqrt{\smash[b]{s_{\footnotesize{NN}}}} \geq 5$
GeV from \cite{Okorokov-AHEP-2015-790646-2015}. In
Figs.~\ref{fig:3.3}, \ref{fig:3.4} solid curves for
$R_{\mbox{\scriptsize{m}}}(\sqrt{\smash[b]{s_{\footnotesize{NN}}}})$
and for
$R_{\mbox{\scriptsize{m}}}^{n}(\sqrt{\smash[b]{s_{\footnotesize{NN}}}})$
respectively are obtained with help of results from fits of BEC
radii by general view of (\ref{eq:3.1}) and dashed curves
correspond to the calculations with fit results for BEC radii for
special case of (\ref{eq:3.1}) at $a_{3}=1.0$. As seen in
Figs.~\ref{fig:3.3}, \ref{fig:3.4} the behavior of smooth curves
with respect to each other as well as to the experimental data are
quite similar for the $R_{\mbox{\scriptsize{m}}}$ and
$R_{\mbox{\scriptsize{m}}}^{n}$. In both cases curves correspond
to experimental points reasonably at intermediate energies $10
\lesssim \sqrt{\smash[b]{s_{\footnotesize{NN}}}} \lesssim 200$ GeV
with excess over experimental points in dip region
$5 \lesssim \sqrt{\smash[b]{s_{\footnotesize{NN}}}} < 10$ GeV,
opposite situation is seen for TeV energies (Figs.~\ref{fig:3.3},
\ref{fig:3.4}). In Figs.~\ref{fig:3.3}, \ref{fig:3.4} solid
curves are close to dashed ones in total energy range
considered especially for $R_{\mbox{\scriptsize{m}}}$ but onset of
the excess of solid curve over dashed one at
$\sqrt{\smash[b]{s_{\footnotesize{NN}}}}
> 1$ TeV for $R_{\mbox{\scriptsize{m}}}$ (Fig.~\ref{fig:3.3}) and
at $\sqrt{\smash[b]{s_{\footnotesize{NN}}}}
> 0.2$ TeV for $R_{\mbox{\scriptsize{m}}}^{n}$ (Fig.~\ref{fig:3.4})
can lead to a noticeable discrepancy at FCC energy.

As expected, the quantitative comparisons of Figs.~\ref{fig:3.1}b-d with Fig.~
\ref{fig:3.2}b for $p+p$ reactions and Fig.~\ref{fig:3.3} with results for
BEC radii $R_{i}$, $i=\mbox{s}, \mbox{o}, \mbox{l}$ in nuclear
collisions \cite{Okorokov-AHEP-2015-790646-2015} show that
$\forall\,i=\mbox{s}, \mbox{o}, \mbox{l}:
R_{\mbox{\scriptsize{m}}} \sim R_{i}$ at qualitative level.

\begin{table*}
\caption{\label{tab:3.2}Estimations for space-time characteristic
of pion source at the LHC and FCC energies}
\begin{ruledtabular}
\begin{tabular}{lcccccc}
\multicolumn{1}{l}{parameter} & \multicolumn{2}{c}{$p+p$, $\sqrt{\smash[b]{s_{\footnotesize{pp}}}}$ (TeV)}&
\multicolumn{2}{c}{$p+\mbox{Pb}$,$\sqrt{\smash[b]{s_{\footnotesize{NN}}}}$ (TeV)} &
\multicolumn{2}{c}{$\mbox{Pb}+\mbox{Pb}$,$\sqrt{\smash[b]{s_{\footnotesize{NN}}}}$ (TeV)} \rule{0pt}{10pt}\\ \cline{2-7}
 & 14 & 100 & 5.02 & 63 & 5.52 & 39 \rule{0pt}{10pt}\\
\hline
\multicolumn{7}{c}{from fits by approach (i.1)} \rule{0pt}{10pt}\\ 
\hline
$\lambda$                       & -- & -- & $0.16 \pm 0.19$ & $0.05 \pm 0.22$ & $0.41 \pm 0.03$ & $0.40 \pm 0.03$ \rule{0pt}{10pt}\\
$R_{\mbox{\scriptsize{s}}}$, fm & $1.8 \pm 0.3$   & $2.3 \pm 0.4$     & $3.9 \pm 2.9$       & $5 \pm 4$   & $6.8 \pm 1.9$ & $8 \pm 3$     \\
$R_{\mbox{\scriptsize{o}}}$, fm & $1.4 \pm 0.6$   & $1.6 \pm 0.7$     & $3.5 \pm 2.6$   & $3.6 \pm 2.7$   & $6.3 \pm 1.0$ & $6.4 \pm 1.0$     \\
$R_{\mbox{\scriptsize{l}}}$, fm & $1.85 \pm 0.07$ & $2.12 \pm 0.11$   & $4 \pm 3$       & $5 \pm 3$       & $7.6 \pm 1.5$ & $8.1 \pm 1.6$ \\
$R_{\mbox{\scriptsize{o}}}/R_{\mbox{\scriptsize{s}}}$ & $0.7 \pm 0.4$ & $0.9 \pm 1.0$   & $0.7 \pm 0.7$   & $0.7 \pm 0.7$ & $0.9 \pm 0.3$ & $0.8 \pm 0.3$ \\
$R_{\mbox{\scriptsize{m}}}$, fm & $1.67 \pm 0.27$   & $2.0 \pm 0.3$     & $3.9 \pm 1.7$   & $4.4 \pm 1.9$   & $6.9 \pm 0.9$ & $7.6 \pm 1.2$    \\
$\delta$, fm$^{2}$              & $-1.5 \pm 2.0$  & $-2.7 \pm 2.9$    & $-3 \pm 29$    & $-20 \pm 50$      & $-6 \pm 29$ & $-30 \pm 60$     \\
$\Delta \tau$, fm/$c$           & $1.5 \pm 1.0$   & $2.0 \pm 1.1$     & $2 \pm 11$       & $5 \pm 7$      & $3 \pm 7$ & $7 \pm 7$     \\
$V$, fm$^{3}$                   & $(1.0 \pm 0.3) \times 10^{2}$ & $(1.8 \pm 0.5) \times 10^{2}$ & $(1.0 \pm 1.7) \times 10^{3}$ & $(2 \pm 3) \times 10^{3}$  & $(6 \pm 3) \times 10^{3}$ & $(9 \pm 8) \times 10^{3}$     \\
\hline
\multicolumn{7}{c}{from fits by approach (i.2)} \rule{0pt}{10pt}\\ 
\hline
$\lambda$                       & $0.42 \pm 0.11$ & $0.42 \pm 0.13$ & $0.097 \pm 0.004$ & --              & $0.362 \pm 0.009$ & $0.315 \pm 0.011$ \rule{0pt}{10pt}\\
$R_{\mbox{\scriptsize{s}}}$, fm & $1.7 \pm 0.7$   & $2.0 \pm 0.8$     & $3.3 \pm 2.4$       & $3.5 \pm 2.6$   & $5.79 \pm 0.10$ & $6.11 \pm 0.12$     \\
$R_{\mbox{\scriptsize{o}}}$, fm & $1.4 \pm 0.7$   & $1.6 \pm 0.9$     & $3.6 \pm 2.6$   & $3.7 \pm 2.7$   & $6.49 \pm 0.12$ & $6.74 \pm 0.15$     \\
$R_{\mbox{\scriptsize{l}}}$, fm & $1.82 \pm 0.17$ & $1.96 \pm 0.20$   & $5 \pm 3$       & $5 \pm 4$       & $8.20 \pm 0.16$ & $9.04 \pm 0.19$ \\
$R_{\mbox{\scriptsize{o}}}/R_{\mbox{\scriptsize{s}}}$ & $0.8 \pm 0.5$ & $0.8 \pm 0.5$   & $1.1 \pm 1.1$   & $1.1 \pm 1.1$ & $1.12 \pm 0.03$ & $1.10 \pm 0.03$ \\
$R_{\mbox{\scriptsize{m}}}$, fm & $1.6 \pm 0.4$   & $1.9 \pm 0.4$     & $3.7 \pm 1.6$   & $4.0 \pm 1.7$   & $6.75 \pm 0.07$ & $7.19 \pm 0.09$    \\
$\delta$, fm$^{2}$              & $-1 \pm 3$  & $-2 \pm 4$    & $2 \pm 25$    & $1 \pm 27$      & $8.7 \pm 2.0$ & $8.2 \pm 2.4$     \\
$\Delta \tau$, fm/$c$           & $1.2 \pm 1.9$   & $1.5 \pm 2.1$     & $2 \pm 11$       & $1 \pm 14$      & $3.6 \pm 0.4$ & $3.5 \pm 0.5$     \\
$V$, fm$^{3}$                   & $80 \pm 60$ & $(1.3 \pm 1.0) \times 10^{2}$ & $(0.8 \pm 1.3) \times 10^{3}$ & $(1.0 \pm 1.6) \times 10^{3}$  & $(4.32 \pm 0.17) \times 10^{3}$ & $(5.31 \pm 0.23) \times 10^{3}$     \\
\end{tabular}
\end{ruledtabular}
\end{table*}

Predictions for values of the BEC observables from sets
$\mathcal{G}_{1,2}$ are obtained for suggested types of
collisions and energies of the LHC and FCC project \cite{Schaumann-arXiv-1503.09107,Armesto-NPA-931-1163-2014,*Dainese-arXiv-1605.01389,Armesto-arXiv-1601.02963} based on the fit
results for the main BEC parameters discussed above and in
\cite{Okorokov-AHEP-2015-790646-2015}. Estimations are shown in
Table~\ref{tab:3.2} for fits by function (\ref{eq:3.1}) and its specific case at
$a_{3}=1.0$ with inclusion of statistical
errors of experimental points, the first column for each type of collisions corresponds to the nominal LHC energy and the second column -- to the energy of FCC project. One notes the fit by constant predicts $\lambda=0.422 \pm 0.004$ for the strength of correlations in $p+p$ collisions for both the LHC and FCC energies. Values for all additional BEC parameters $\mathcal{G}_{2}^{j}$, $j=1-4$ are calculated with help of its definitions (\ref{eq:2.3}) -- (\ref{eq:2.5}a) and estimations for BEC radii at some energy. The pion emission duration is derived from $|\delta|$ and kinematic regime for pion pairs under study as well as for $p+p$ at lower energies. Results for asymmetric collisions $p+\mbox{Pb}$ are
obtained with help of fit results for scaled BEC parameters
\cite{Okorokov-AHEP-2015-790646-2015} and rough estimation
$\langle R_{p\mbox{\scriptsize{Pb}}}\rangle=(4 \pm 3)$ fm. In the case of $\mbox{Pb}+\mbox{Pb}$ collisions the results
for main BEC parameters $\mathcal{G}_{1}^{i}$, $i=1-4$ as well as for $\mathcal{G}_{2}^{j}$, $j=1,4$ and emission duration are from
\cite{Okorokov-AHEP-2015-790646-2015}, furthermore the brief discussion of estimations for these BEC parameters at the LHC and FCC energies can be also found in the previous studies \cite{Okorokov-AHEP-2015-790646-2015,Okorokov-arXiv-1504.08336}. As seen from Table~\ref{tab:3.2} all BEC parameters have values coincided with each other for two approaches (i.1) and (i.2) within errors for corresponding collision energies and types of strong interaction processes. In general estimations for BEC parameters calculated with approach (i.1) do not change from the LHC up to FCC energies within large uncertainties for all collisions under consideration. Proton-proton and nucleus-nucleus collisions are characterized by similar strength of correlations for approach (i.2) for the LHC and FCC energies, the large uncertainty for $p+\mbox{Pb}$ allows only the qualitative conclusion that $\lambda$ is somewhat smaller for this collision type than that for $\mbox{Pb}+\mbox{Pb}$ in the case of general view of (\ref{eq:3.1}) and for $p+p$, $\mbox{Pb}+\mbox{Pb}$ within the framework of the approach (i.2). The $\lambda$ is quite constant for $p+p$ collisions but shows noticeable decrease for heavy ion collisions at increasing of $\sqrt{\smash[b]{s_{\footnotesize{NN}}}}$ in the energy range LHC -- FCC for fit by specific case of (\ref{eq:3.1}) at $a_{3}=1.0$. For $p+\mbox{Pb}$ the estimation of $\lambda$ obtained with approach (i.1) at $\sqrt{\smash[b]{s_{\footnotesize{NN}}}}=5.02$ TeV agrees rather well with experimental result \cite{Adam-PRC-91-034906-2015} while the approach (i.2) underpredicts the strength of correlations at the LHC energy. Furthemore estimations for BEC radii of pion source produced in $p+\mbox{Pb}$ collisions at $\sqrt{\smash[b]{s_{\footnotesize{NN}}}}=5.02$ TeV (Table~\ref{tab:3.2}) agree with experimental results \cite{Adam-PRC-91-034906-2015,Koehler-arXiv-1601.05632-2016} within large errors. The space scales are about 2 fm in $p+p$, about 4 -- 5 fm in $p+\mbox{Pb}$ and 6 -- 9 fm in $\mbox{Pb}+\mbox{Pb}$ collisions for pion source at FCC energies. In the case of estimations for BEC radii obtained with general view of (\ref{eq:3.1}) large uncertainties do not allow the definite conclusion and one can see qualitative indication only that $R_{\mbox{\scriptsize{o}}}$ is some smaller than other radii in all collision types for both the LHC and FCC energies. For estimations based on the special case of (\ref{eq:3.1}) with $a_{3}=1.0$ there is noticeable increase of all BEC radii for transition from the LHC to FCC energy in heavy ion collision (Table~\ref{tab:3.2}). As consequence the $R_{\mbox{\scriptsize{m}}}$ and $V$ is larger at $\sqrt{\smash[b]{s_{\footnotesize{NN}}}}=39$ TeV than that at the LHC energy $\sqrt{\smash[b]{s_{\footnotesize{NN}}}}=5.52$ TeV. Spread of values of BEC radii leads to significant uncertainties for estimations of additional space-time parameters especially for extremely asymmetric $p+\mbox{Pb}$ collisions for which the large error for $\langle R_{p\mbox{\scriptsize{Pb}}}\rangle$ increases greatly uncertainties for BEC quantities in Table~\ref{tab:3.2}.
Consequently the volume of the pion source can be roughly estimated as about $200$ fm$^{3}$ in $p+p$, $2000$ fm$^{3}$ in $p+\mbox{Pb}$ and $10^{4}$ fm$^{3}$ in $\mbox{Pb}+\mbox{Pb}$ collisions at FCC energies in comparison with $100$ fm$^{3}$ in $p+p$, $1000$ fm$^{3}$ in $p+\mbox{Pb}$ and $6000$ fm$^{3}$ in $\mbox{Pb}+\mbox{Pb}$ at the nominal LHC energies. These estimations indicate the consistent growth of $V$ for transition from the small system collisions to the $\mbox{Pb}+\mbox{Pb}$. For approach (i.1) estimations for all parameters $\mathcal{G}_{2}^{j}$, $i=j-4$ and for emission duration do not depend on energy in the range LHC -- FCC for all types of collisions within errors. This conclusion is also valid for approach (i.2) with exception of the $R_{\mbox{\scriptsize{m}}}$ and $V$ for $\mbox{Pb}+\mbox{Pb}$ discussed above. It should be noted that weak change of main BEC parameters $\mathcal{G}_{1}^{i}$, $i=1-4$ is qualitatively expected for energy domain from the LHC up to FCC because of general trends of available experimental points and consequent slow logarithmic increase with collision energy for smooth analytic functions used in the present analysis as well as in \cite{Okorokov-AHEP-2015-790646-2015,Okorokov-arXiv-1504.08336}.

\section{Pion laser at FCC energies}\label{sec.4}
\begin{figure}[b]
\resizebox{0.46\textwidth}{!}{%
\includegraphics[width=7.5cm,height=6.8cm]{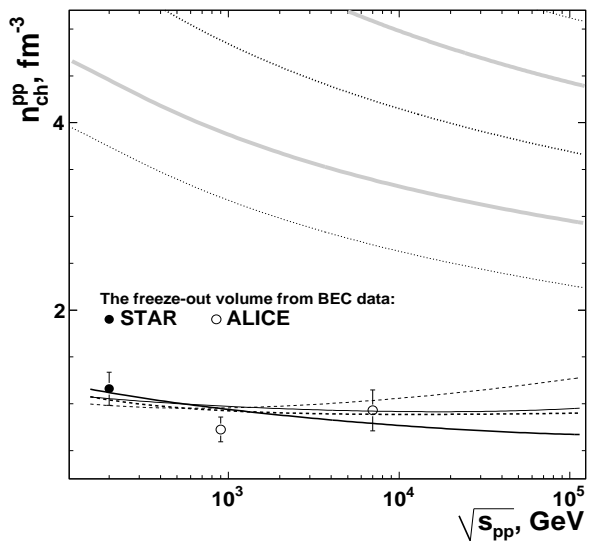}}\caption{\label{fig:4.1} Energy
dependence of estimations for charged particle density and for
critical one in $p+p$ collisions. Points are calculated with help of the hybrid function for $N_{ch}$ \cite{Sarkisyan-PR-D93-054046-2016} and the experimental estimations for $V$ and uncertainties for points are propagated from statistical errors of measurements and fits used.
Solid lines correspond to the hybrid
approximation of $N_{ch}$ \cite{Sarkisyan-PR-D93-054046-2016} and
dashed lines are for 3NLO pQCD equation
\cite{Dremin-PR-349-301-2001} while thick lines show results with
$V$ calculated with the fits of BEC radii by (\ref{eq:3.1}) in the
case (ii.1) and thin lines -- with the fits by specific case
$R_{i} \propto \ln\varepsilon$, $i=\mbox{s}, \mbox{l}$. Critical
charged particle density is shown by dotted line with its
statistical uncertainty levels represented by thin dotted lines.
The heavy grey lines correspond to the systematic $\pm 1$ s.d. of
$n_{\mbox{\scriptsize{ch}}}$ calculated by varying of $\eta$ on
$\pm 0.05$.}
\end{figure}

\begin{figure}[b]
\resizebox{0.46\textwidth}{!}{%
\includegraphics[width=7.5cm,height=6.8cm]{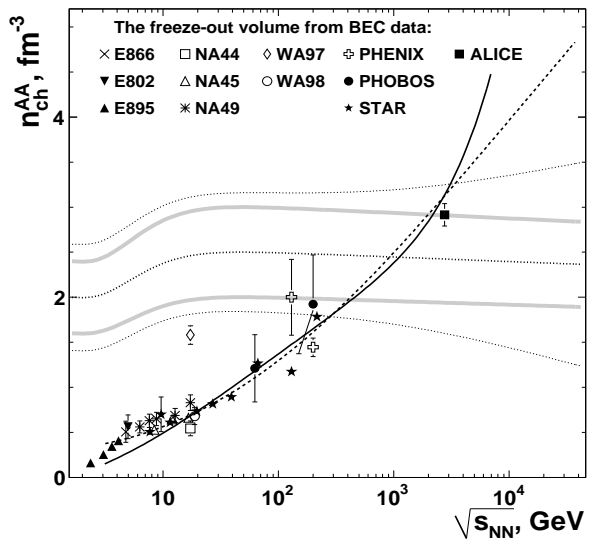}
}\caption{\label{fig:4.2} Energy dependence of estimations for
charged particle density and for critical one in
$\mbox{A}+\mbox{A}$ collisions. Points are calculated with help of
the hybrid function for $N_{ch}$ \cite{Abbas-PLB-726-610-2013} and
the experimental estimations for $V$ and uncertainties for points
are propagated from statistical errors of measurements and fits
used. Solid line corresponds to the hybrid approximation of
$N_{ch}$ \cite{Sarkisyan-PR-D93-054046-2016} and dashed line is
for parametrization of total charged multiplicity from
\cite{Abbas-PLB-726-610-2013}. Critical charged particle density
is shown by dotted line with its statistical uncertainty levels
represented by thin dotted lines. The heavy grey lines correspond
to the systematic $\pm 1$ s.d. of $n_{\mbox{\scriptsize{ch}}}$
calculated by varying of $\eta$ on $\pm 0.05$.}
\end{figure}

Results shown above for space-time extent of pion emissions region
allow the study of possibility of Bose\,--\,Einstein condensation
with consequent formation of pion laser in high energy strong
interaction processes. The key quantity is the charged particle
density which is defined as follows:
\begin{equation}
n_{\mbox{\scriptsize{ch}}}=N_{\mbox{\scriptsize{ch}}}/V,
\label{eq:4.1}
\end{equation}
where $N_{\mbox{\scriptsize{ch}}}$ is the total charged particle
multiplicity and $V$ is the source volume at freeze-out stage. The
critical density
$n_{\mbox{\scriptsize{ch}}}^{\mbox{\scriptsize{c}}}$ can be
calculated with help of (\ref{eq:4.1}) and transition to the
critical total multiplicity $N_{\mbox{\scriptsize{ch}}} \to
N_{\mbox{\scriptsize{ch}}}^{\mbox{\scriptsize{c}}}$. The last
multiplicity parameter was derived for 1D thermal Gaussian
distribution in \cite{Pratt-PLB-301-159-1993}. Within the 3D Gaussian parametrization for
source the following relation is suggested
\begin{equation}
N_{\mbox{\scriptsize{ch}}}^{\mbox{\scriptsize{c}}}=\eta^{-1}\bigl[(\Delta
R_{\mbox{\scriptsize{m}}})^{2}+
\sqrt{\smash[b]{(p_{0}R_{\mbox{\scriptsize{m}}})^{2}+0.25}}\bigr]^{3/2}.\label{eq:4.2}
\end{equation}
Here $\eta=0.25$ is the fraction of the pions to be emitted from a
static Gaussian source, $\Delta=0.25$ GeV/$c$ is a momentum spread
and $p_{0}/T=p^{\,2}/2\Delta^{2}$ \cite{Pratt-PLB-301-159-1993},
$T$ -- source temperature supposed equal to the parameter value at
chemical freeze-out, it is suggested $p \approx 3T$
\cite{Shuryak-book-2004}. It should be emphasized that there are
no qualitative studies of
$T(\sqrt{\smash[b]{s_{\footnotesize{pp}}}})$ for $p+p$ collisions
so far but set of results for mean multiplicity and the pseudorapidity density at midrapidity for charged particles in various collisions \cite{Sarkisyan-arXiv-hep-ph-0410324-2004,*Sarkisyan-AIP-828-35-2006,*Sarkisyan-EPJ-C70-533-2010,Sarkisyan-PR-D93-054046-2016} as well as recent results for deconfinement in small system
\cite{Abelev-PRC-79-034909-2009,Gutay-IJMPE-24-1550101-2015} indicate remarkable similarity
of both the bulk and the thermodynamic properties of strongly interacting matter created
in high energy $p+p$ / $\bar{p}+p$ and $\mbox{A}+\mbox{A}$ collisions.
Therefore the hypothesis is suggested for similar energy
dependence of $T$ in both the $p+p$ and the $\mbox{A}+\mbox{A}$
interactions with taking into account
$\sqrt{\smash[b]{s_{\footnotesize{pp}}}} \simeq
3\sqrt{\smash[b]{s_{\footnotesize{NN}}}}$
\cite{Sarkisyan-arXiv-hep-ph-0410324-2004,*Sarkisyan-AIP-828-35-2006,*Sarkisyan-EPJ-C70-533-2010,Sarkisyan-PR-D93-054046-2016} and consequently the analytic
energy dependence of $T$ from
\cite{Cleymans-JPhysG-32-S165-2006,*Cleymans-PRC-73-034905-2006}
is used for all types of strong interaction processes
considered in this Section. Thus appropriate analytic function is derived for energy dependence of $N_{\mbox{\scriptsize{ch}}}^{\mbox{\scriptsize{c}}}$. Also $N_{\mbox{\scriptsize{ch}}}$ vs collision energy in (\ref{eq:4.1}) is defined by some smooth approximations which are specified below for $p+p$ and $\mbox{A}+\mbox{A}$ collisions. Experimental estimations for $V$ available in the Fig.~\ref{fig:3.2}e for $p+p$ and in \cite{Okorokov-AHEP-2015-790646-2015} for $\mbox{A}+\mbox{A}$ collisions can be used for calculations of $n_{\mbox{\scriptsize{ch}}}$ at certain energies. The results of such calculations are called experimental points and marked by symbols in Figures \ref{fig:4.1} and \ref{fig:4.2} below in the sense that BEC measurements are used in these cases.
As seen from Table~\ref{tab:3.1} for $p+p$ and from
\cite{Okorokov-AHEP-2015-790646-2015} for heavy ion collisions the
smooth energy dependence of source volume can be calculated with
help of the fits of BEC radii by (ii.1) general function
(\ref{eq:3.1}) as well as (ii.2) by specific case $R_{i} \propto
\ln\varepsilon$, $i=\mbox{s}, \mbox{l}$. The relation
(\ref{eq:2.5}a) for source volume is used in (\ref{eq:4.1}) for
experimental estimations as well as for calculations of smooth
energy dependence of $n_{\mbox{\scriptsize{ch}}}$ in both the
$p+p$ and the $\mbox{A}+\mbox{A}$ collisions. On the other hand it
seems reasonable to use the equation (\ref{eq:2.5}b) for
estimation of critical charged density because the relation
(\ref{eq:4.2}) with $R_{\mbox{\scriptsize{m}}}$ is only available.
Taking into account the qualitative relation between
$R_{\mbox{\scriptsize{m}}}$ and 3D BEC radii shown above one
expects that the difference of numerical factors for two
definition of $V$ (\ref{eq:2.5}) provides additional
uncertainty $\sim 3\sqrt{\pi/2}$ in the ratio
$n_{\mbox{\scriptsize{ch}}}/n_{\mbox{\scriptsize{ch}}}^{\mbox{\scriptsize{c}}}$.
Therefore theoretical investigations are essential for
quantitative account for geometry of source and decrease of
uncertainty due to calculation of volume of emission region.
As it follows from (\ref{eq:2.dop}) and discussion in the Sec.~\ref{sec.2} the relation (\ref{eq:4.1}) defines the upper boundary for true value of charged particle density
\begin{equation}
n_{\mbox{\scriptsize{ch}}}=\sup n_{\mbox{\scriptsize{ch}}}^{\mbox{\scriptsize{tr}}}
\label{eq:4.dop}
\end{equation}
strictly speaking and possibly the discrepancy between estimation from (\ref{eq:4.1}) and $n_{\mbox{\scriptsize{ch}}}^{\mbox{\scriptsize{tr}}}$ will be increased with growth of collision energy due to intensification of longitudinal and radial collective flows. During the recent years the collectivity in small colliding systems are under intensive theoretical \cite{Li-MPLA-27-1230018-2012,*Bozek-PRC-85-014911-2012,*Bozek-PLB-718-1557-2013,*Bjorken-PLB-726-344-2013,*Dusling-PRD-87-051502-2013,*Dusling-PRD-87-054014-2013,*Schenke-PRL-113-102301-2014,*Dusling-IJMPE-25-1630002-2016} and experimental studies \cite{Khachatryan-JHEP-0910-091-2010,*Aad-PRL-116-172301-2016,*Khachatryan-PRL-116-172302-2016} but the results obtained already mean that the statement above is valid for $p+p$ interactions as well as for nucleus-nucleus collisions
in TeV-energy domain at least. In general one can assume the weaker dependence of $n_{\mbox{\scriptsize{ch}}}^{\mbox{\scriptsize{c}}}$ on space-time parameters of particle source than that for $n_{\mbox{\scriptsize{ch}}}$ due to BEC radius in (\ref{eq:4.2}). The uncertainties of $N_{\mbox{\scriptsize{ch}}}$ calculated by standard way from errors of fit parameters are attributed as statistical errors. The
statistical uncertainties for
$N_{\mbox{\scriptsize{ch}}}^{\mbox{\scriptsize{c}}}$ are estimated
by standard way at assigned relative error $\delta \Delta=0.05$ and with taking into account the errors for fits of $R_{\mbox{\scriptsize{m}}}$ and $T$ used
while systematic uncertainties are deduced by varying of $\eta$
within the range $\eta=0.20-0.30$ only. The statistical errors of $V$ propagated from corresponding uncertainties of BEC radii are only taken into account below.

Total charged multiplicity
$N_{\mbox{\scriptsize{ch}}}^{\mbox{\scriptsize{pp}}}$ is
calculated within various approaches
\cite{Sarkisyan-PR-D93-054046-2016,Dremin-PR-349-301-2001,Abbas-PLB-726-610-2013}.
In Fig.~\ref{fig:4.1} energy dependence is shown for
$n_{\mbox{\scriptsize{ch}}}^{\mbox{\scriptsize{pp}}}$ as well as
for critical particle density. It should be noted that in the case of $p+p$ collisions minimum-bias events are used in the BEC analyses \cite{Aggarwal-PRC-83-064905-2011,Aamodt-PRD-84-112004-2011} and these events correspond to the non-singly diffractive (NSD) $p+p$ collisions at $\sqrt{\smash[b]{s_{\footnotesize{pp}}}}=0.2$ TeV \cite{Abelev-PRC-79-034909-2009} as well as at the LHC energies \cite{Aamodt-EPJC-68-89-2010,Adam-arXiv-1509.07541-nuc-ex-2015}. Thus experimental points are obtained
for $N_{\mbox{\scriptsize{ch}}}^{\mbox{\scriptsize{pp}}}$
calculated with hybrid approximation
\cite{Sarkisyan-PR-D93-054046-2016}. The thick solid and dashed
lines correspond to the source volume within approach (ii.1) and
thin lines -- to the $V$ from case (ii.2). In Fig.~\ref{fig:4.1}
results for $n_{\mbox{\scriptsize{ch}}}^{\mbox{\scriptsize{pp}}}$
are shown by solid lines for hybrid approximation of total charged
multiplicity \cite{Sarkisyan-PR-D93-054046-2016} and dashed lines
are deduced with
$N_{\mbox{\scriptsize{ch}}}^{\mbox{\scriptsize{pp}}}$ within 3NLO
pQCD approach \cite{Dremin-PR-349-301-2001} at
$\sqrt{\smash[b]{s_{\footnotesize{pp}}}} \simeq
0.35\sqrt{\smash[b]{s_{ee}}}$
\cite{Sarkisyan-arXiv-hep-ph-0410324-2004,*Sarkisyan-AIP-828-35-2006,*Sarkisyan-EPJ-C70-533-2010,Grosse-Oetringhaus-JPGNPP-37-083001-2010} and parameters
from \cite{Heister-EPJC-35-457-2004} for number of colors
$N_{c}=3$. Available experimental estimations show almost constant
$n_{\mbox{\scriptsize{ch}}}^{\mbox{\scriptsize{pp}}} \simeq 1$.
Smooth curves agree with experimental points reasonably for any
approximation of
$N_{\mbox{\scriptsize{ch}}}^{\mbox{\scriptsize{pp}}}$ and $V$. As
seen the differences between various approaches for each of the
two parameters are small up to the LHC energy and increase for FCC
noticeably. The dependence of critical particle density
$n_{\mbox{\scriptsize{ch}}}^{\mbox{\scriptsize{c,pp}}}(\sqrt{\smash[b]{s_{\footnotesize{pp}}}})$
shown by dotted line decreases with energy. The statistical $\pm
1$ s.d. band limits are drawn by thin dotted line while this large
uncertainty is mostly dominated by the precision of BEC parameters
of emission region. The systematic $\pm 1$ s.d. boundaries are
shown by heavy grey lines. As seen from Fig.~\ref{fig:4.1} the
$n_{\mbox{\scriptsize{ch}}}$ is smaller than its critical value in
$p+p$ collisions up to FCC energy
$\sqrt{\smash[b]{s_{\footnotesize{pp}}}}=100$ TeV for any
approaches for total charged multiplicities and $V$ under study.
This conclusion is valid even with taking into account large
statistical uncertainties. Thus one can not expect the kind of
lasing behavior for secondary pions in $p+p$ collisions within the
present approach.

One can note the coincidence between the experimental values of $N_{\mbox{\scriptsize{ch}}}$ in $p+p$ and $\bar{p}+p$ collisions at energies 0.2 and 0.9 TeV and general smooth energy dependence of $N_{\mbox{\scriptsize{ch}}}$ for NSD events in the interactions under discussion \cite{Adam-arXiv-1509.07541-nuc-ex-2015}. Also reasonable agreement is observed for pseudorapidity density $dN_{\mbox{\scriptsize{ch}}}/d\eta$ measured in $p+p$ \cite{Aamodt-EPJC-68-89-2010,Adam-arXiv-1509.07541-nuc-ex-2015,Nouicer-JPGNPP-30-S1133-2004} and $\bar{p}+p$ \cite{Alner-ZPC-33-1-1986,*Ansorge-ZPC-43-357-1989} interactions at energies indicated above. But there are no 1D BEC analyses with Gaussian model in $\bar{p}+p$ collisions at $\sqrt{\smash[b]{s_{\footnotesize{\bar{p}p}}}}=0.2$ and 0.9 TeV. Furthermore the quantitative comparison of BEC results from $p+p$ to those from $\bar{p}+p$ collisions is difficult for 1D case due to limited ensemble of experimental results in the last case \cite{Okorokov-arXiv-1605.02927} and noticeable difference between collision energies in $p+p$ and $\bar{p}+p$ for available BEC measurements; there is no BEC analysis with 3D Gaussian model for $\bar{p}+p$ so far. On qualitative level close values for multiplicity quantities can be expected in $p+p$ and $\bar{p}+p$ collisions in particular at energies about 2 TeV while the 1D BEC Gaussian radius for pion source in $\bar{p}+p$ at $\sqrt{\smash[b]{s_{\footnotesize{\bar{p}p}}}}=1.8$ and 1.96 TeV \cite{Alexopoulos-PRD-48-1931-1993,*Lovas-PhDThesis-2008} is significantly larger than that from $p+p$ at $\sqrt{\smash[b]{s_{\footnotesize{pp}}}}=2.36$ TeV \cite{Khachatryan-PRL-105-032001-2010}. Therefore the $n_{\mbox{\scriptsize{ch}}}$ is expected smaller in $\bar{p}+p$ than that in $p+p$ collisions at least in TeV-energy domain at close values of critical quantity due to its weaker dependence on space-time extent of particle source. Thus the pion lasers seem impossible in high-energy $\bar{p}+p$ collisions within the rough assumptions.

As discussed above the estimations of space-time extent of pion
source are characterized by large uncertainties moreover
development of equation for critical parameters for
multidimensional (3D) case seems important for improvement of
precision of studies and for more certain conclusions. The future
quantitative experimental and theoretical investigations are
essential for verification of the results shown above and
possibility of novel coherent effects in different types of collisions in
high energy domain.

Total charged multiplicity
$N_{\mbox{\scriptsize{ch}}}^{\mbox{\scriptsize{AA}}}$ is
calculated with hybrid equations
\cite{Sarkisyan-PR-D93-054046-2016,Abbas-PLB-726-610-2013}. Fig.~\ref{fig:4.2} demonstrates energy dependence for both the
$n_{\mbox{\scriptsize{ch}}}^{\mbox{\scriptsize{AA}}}$ and the
critical particle density where smooth approximation for
$n_{\mbox{\scriptsize{ch}}}^{\mbox{\scriptsize{AA}}}$ is shown by
solid line for hybrid approximation of total charged multiplicity
\cite{Sarkisyan-PR-D93-054046-2016} and by dashed line for
$N_{\mbox{\scriptsize{ch}}}^{\mbox{\scriptsize{AA}}}$ from
\cite{Abbas-PLB-726-610-2013}, and experimental points are obtained
for $N_{\mbox{\scriptsize{ch}}}^{\mbox{\scriptsize{AA}}}$
calculated with equation from \cite{Abbas-PLB-726-610-2013}. Both
the curves for
$n_{\mbox{\scriptsize{ch}}}^{\mbox{\scriptsize{AA}}}$ and
experimental points are deduced with mean number of participant
$\langle N_{\mbox{\scriptsize{part}}}\rangle=382.8 \pm 3.1$ which corresponds to
the 0--5\% central $\mbox{Pb}+\mbox{Pb}$ collisions
\cite{Abbas-PLB-726-610-2013}. This simplest approach seems
reasonable because heavy ion collisions are only considered in
Fig.~\ref{fig:4.2}. The source volume is calculated within
approach (ii.1) and results for BEC radii in nuclear collisions
\cite{Okorokov-AHEP-2015-790646-2015}. Experimental points for
$n_{\mbox{\scriptsize{ch}}}^{\mbox{\scriptsize{AA}}}$ increase
with $\sqrt{\smash[b]{s_{\footnotesize{NN}}}}$ and agree
reasonably with smooth curves for both parameterizations of total
charged multiplicity under consideration. Comparison between
particle densities in $p+p$ (Fig.~\ref{fig:4.1}) and
$\mbox{A}+\mbox{A}$ (Fig.~\ref{fig:4.2}) strong interaction
processes indicates the enhancement of
$n_{\mbox{\scriptsize{ch}}}^{\mbox{\scriptsize{AA}}}$ over
$n_{\mbox{\scriptsize{ch}}}^{\mbox{\scriptsize{pp}}}$ starting
with RHIC energy 200 GeV per nucleon-nucleon pair furthermore this
enhancement increases with growth of collision energy. The
critical particle density
$n_{\mbox{\scriptsize{ch}}}^{\mbox{\scriptsize{c,AA}}}$ shown by
dotted line depends weakly on
$\sqrt{\smash[b]{s_{\footnotesize{NN}}}}$ in nuclear collisions.
In Fig.~\ref{fig:4.2} the line types for statistical and
systematic $\pm 1$ s.d. band limits are the same as well as for
corresponding smooth curves for $p+p$ collisions
(Fig.~\ref{fig:4.1}). The statistical uncertainty driven
by the precision of BEC parameters of emission region increases
noticeably for multi-TeV region
$\sqrt{\smash[b]{s_{\footnotesize{NN}}}} \gtrsim 10$ TeV in $\mbox{A}+\mbox{A}$ interactions. The
situation changes dramatically with transition from $p+p$ to
nuclear collisions at high energies. As seen from
Fig.~\ref{fig:4.2} the following relation
$n_{\mbox{\scriptsize{ch}}}^{\mbox{\scriptsize{AA}}} \approx
n_{\mbox{\scriptsize{ch}}}^{\mbox{\scriptsize{c,AA}}}$ is valid at
RHIC and the LHC energies within wide uncertainty band for
critical value of charged particle density. Furthermore there is
clear enhancement of smooth curves for
$n_{\mbox{\scriptsize{ch}}}^{\mbox{\scriptsize{AA}}}$ over
$n_{\mbox{\scriptsize{ch}}}^{\mbox{\scriptsize{c,AA}}}$ at
$\sqrt{\smash[b]{s_{\footnotesize{NN}}}} \gtrsim 10$ TeV with
taking into account large statistical uncertainty for
$n_{\mbox{\scriptsize{ch}}}^{\mbox{\scriptsize{c,AA}}}(\sqrt{\smash[b]{s_{\footnotesize{NN}}}})$.
Thus one can expect the appearance of novel effects dominated by
Bose\,--\,Einstein condensation in nucleus-nucleus collisions at
FCC energy. In particular, Fig.~\ref{fig:4.2} indicates the
possibility for pion laser effect in heavy ion collisions at
$\sqrt{\smash[b]{s_{\footnotesize{NN}}}} \gtrsim 10$ TeV within
the approach under study.

With theory point of view the conception of the pion laser was intensively studied within the framework of the model of independent factorized sources \cite{Pratt-PLB-301-159-1993,Pratt-PRC-50-469-1994,Csorgo-PRL-80-916-1998,*Csorgo-HIP-9-161-1999,*Zimanyi-HIP-9-241-1999}
as well as in the model of disoriented chiral condensate (DCC) decay \cite{Akkelin-NPA-661-613c-1999}. On the other hand possible experimental signatures of Bose\,--\,Einstein condensation, in particular, the pion laser effect in heavy ion collisions at FCC energy should be the subject of future detailed investigations. Here one notes the following experimental signatures of Bose\,--\,Einstein condensation. In general, one can expects enhancement of high-multiplicity events \cite{Pratt-PLB-301-159-1993,Pratt-PRC-50-469-1994} and the decrease of chaoticity parameter derived from two-particle BEC analysis due to amplification of coherent particle production \cite{Chao-JPGNPP-21-847-1995}. The effects of multi-boson symmetrization regarding isospin fluctuations can manifest itself through enhancement of the events with anomalous isospin imbalance like CENTAURO events in high-energy cosmic ray \cite{Pratt-PLB-301-159-1993,Pratt-PRC-50-469-1994}. The shrinkage of the BEC radius is the more specific prediction within the model of the DCC decay when the Bose\,--\,Einstein condensation takes place \cite{Akkelin-NPA-661-613c-1999}. This effect potentially represents one of the most pronounced features of the pion laser, because the available experimental BEC radii show smooth increase with collision energy both in the $p+p$ interactions (Fig.~\ref{fig:3.1}) and the heavy ion collisions \cite{Okorokov-AHEP-2015-790646-2015,Okorokov-arXiv-1504.08336}.

\section{Study of correlation peak shape\label{sec.5}}
The accelerator parameters within the FCC project \cite{Schaumann-arXiv-1503.09107,Armesto-NPA-931-1163-2014,*Dainese-arXiv-1605.01389,Armesto-arXiv-1601.02963} open the new possibility for detailed study of peak structure for two-particle BEC correlation function. The peak of CF is described by $\mathbf{K}_{2}^{\mbox{\scriptsize{ph}}}({\bf A})$.
In general there is rich class of
random processes with additive stochastic variables for which
(i.e. for these processes) there are finite distributions but the
Central Limit Theorem (CLT) in the traditional (Gaussian)
formulation is not valid. The class of random processes under
consideration are characterized by large fluctuations, power-law
behavior of distributions in the range of large absolute values of
random variables, non-analytic behavior of characteristic function
of the probability distribution for small values of its arguments
\cite{Csorgo-EPJ-C36-67-2004}. In mathematical statistics and
probability theory the class of such distributions are called as
stable (on L\'{e}vy) distributions\footnote{In literature for
physics and mathematics the multidimensional distributions
included in the class are called as L\'{e}vy\,--\,Feldheim
distributions.} \cite{Feller-book-1967,*Zolotarev-book-1983}. The general stable
distribution is described by four parameters: an index of
stability (or L\'{e}vy index) $\alpha \in (0,2]$, a parameters of skewness $\beta$, scale $\gamma$ and location
$\delta$. These distributions satisfy requirements
of generalized Central Limit Theorem (gCLT) and
self-similarity\footnote{The applications of stable distributions
in the physics of fundamental interactions and, in particular, for
correlation femtoscopy are described, for example, in
\cite{Okorokov-UchPosob-2009}.}. Therefore the detail
investigation of the shape of correlation peak has to do with
verification of hypothesis of possible self-affine fractal-like
geometry of emission region. At present the study of
L\'{e}vy\,--\,Feldheim distributions is the advanced region of
mathematics but the specific case of central-symmetrical stable
distributions is known in more detail
\cite{Samorodnitzky-book-1994}. Just this subclass of stable
distributions is most important on the point of view of investigation for BEC. In this case the application of subset of
non-isotropic central-symmetrical L\'{e}vy\,--\,Feldheim
distributions \cite{Uchaikin-ZETF-124-903-2003} seems reasonable
because the projections of the 3D relative momentum $\vec{q}$ are independent random
variables.

The multidimensional generalized parametrization of n-th order
for CF (\ref{eq:2.1}) can be written as follows \cite{Okorokov-arXiv-1312.4269}:
\begin{widetext}
\begin{subequations}
\begin{equation}
\hspace*{-2.3cm}C_{2}^{\,\mbox{\scriptsize{ph}},n}(q,K)=\xi_{1}(q,K)\left[1+
\xi_{2}(q,K)\mathbf{K}_{2}^{\mbox{\scriptsize{ph}},n}({\bf
A})\right], \label{eq:5.1.a}
\end{equation}
\vspace*{-0.5cm}
\begin{equation}
\mathbf{K}_{2}^{\mbox{\scriptsize{ph}},n}({\bf A})=\mathbf{K}_{2}
^{\mbox{\scriptsize{ph}},0}({\bf A})
\prod\limits_{i=1}^{3}\prod\limits_{j=1}^{3}
\biggl[1+\sum\limits_{m=1}^{n}g_{m}h_{m}
(A_{ij})\biggr],~~\mbox{at}~n \geq 1. \label{eq:5.1.b}
\end{equation}
\label{eq:5.1}
\end{subequations}
\end{widetext}
where $\mathbf{K}_{2}^{\mbox{\scriptsize{ph}},n}$
-- phenomenological parametrization of n-th order for cCF (\ref{eq:2.2}), functions $\xi_{1,2}(q,K)$
take into account formally all corrections on degree of source
chaoticity, final state interactions, etc. The experimental and
theoretical investigations in the field of BEC
allow us to derive some approach for cumulant two-particle
function (\ref{eq:2.2}) in the lowest order. Within the framework
of the subset of non-isotropic central-symmetrical
L\'{e}vy\,--\,Feldheim distributions the most general
parametrization of $\mathbf{K}_{2,\mbox{\scriptsize{L}}}^{\mbox{\scriptsize{ph}},0}$
can be given by
\begin{widetext}
\begin{equation}
\mathbf{K}_{2,\mbox{\scriptsize{L}}}^{\mbox{\scriptsize{ph}},0}({\bf
A})=\prod\limits_{i=1}^{3}\prod\limits_{j=1}^{3}
\mathbf{K}_{2,\mbox{\scriptsize{L}}}^{\mbox{\scriptsize{ph}},0}(A_{ij})=
\exp\biggl(-\sum\limits_{i,j=1}^{3}|A_{ij}|^{\alpha/2}\biggr),~~
\mathbf{K}_{2,\mbox{\scriptsize{L}}}^{\mbox{\scriptsize{ph}},0}(x)=\exp(-|x|^{\alpha/2}).
\label{eq:5.2}
\end{equation}
\end{widetext}
Here were take into account that $x \equiv (q_{i}R_{i})^{2}$,
$i=\mbox{l}, \mbox{o}, \mbox{s}$ for correlation femtoscopy, the
products are on the space components of vectors. The
$\left.\{h_{n}(x)\}\right|_{n=0}^{\infty}$ is the closed system of
orthogonal polynomials in the Hilbert space $\mathcal{H}$:
$\displaystyle \int
dx\mathbf{K}_{2}^{\mbox{\scriptsize{Ф}}}(x)h_{n}(x)h_{m}(x)=\delta_{nm}$,
$g_{n}=\displaystyle \int
dx\mathbf{K}_{2}^{\mbox{\scriptsize{Ф}}}(x)h_{n}(x)$. The system
$\left.\{h_{n}(x)\}\right|_{n=0}^{\infty}$ for exponential weight
function can be derived with the help of the following recurrent
relations $a_{1}h_{1}(x)=(x-b_{0})h_{0}(x)$,
$a_{n+1}h_{n+1}(x)=(x-b_{n})h_{n}(x)-a_{n-1}h_{n-1}(x), n=1,2,...$
\cite{Stahl-book-1992,*Magnus-arXiv-math-9611218,*Levin-book-2001,*Kasuda-JApp-121-13-2003} and moments $\displaystyle
\mu_{n}=\int_{-\infty}^{\infty}dx
x^{n}\exp(-|x|^{\gamma})=2\gamma^{-1}\,\Gamma\bigl(\gamma^{-1}[n+1]\bigr),~n
\geq 0,~\gamma > 0$ \cite{Prudnikov-IntegraliBook1-1981}. Here
$\forall~n \geq 0: b_{n}=\tilde{H}_{n+1}H_{n+1}^{-1}-$
$\tilde{H}_{n}H_{n}^{-1}$; $\forall~n
> 0: a_{n}=H_{n}^{-1}\sqrt{\mathstrut H_{n-1}H_{n+1}}$, and
$H_{n}, \tilde{H}_{n}$ are the following determinants:
$$
H_{n}=
\begin{vmatrix}
\mu_{0}   & \dots & \mu_{n-1} \\
\vdots    &       & \vdots \\
\mu_{n-1} & \dots & \mu_{2n-2} \\
\end{vmatrix},
\tilde{H}_{n}=
\begin{vmatrix}
\mu_{0}   & \dots & \mu_{n-2}  & \mu_{n}\\
\vdots    &       & \vdots     & \vdots \\
\mu_{n-1} & \dots & \mu_{2n-3} & \mu_{2n-1}\\
\end{vmatrix},
$$
$H_{0}=1$ and $\tilde{H}_{0}=0$, the $h_{0}(x)=\mbox{const} > 0$
is defined by normalization which is chosen for system
$\left.\{h_{n}(x)\}\right|_{n=0}^{\infty}$ under consideration.
The clear view of $h_{n}(x)$ can be found, for instance, in \cite{Novak-arXiv-1604.05513}
for few lowest orders $n$ at normalization $h_{0}(x)=1$.
The specific cases $\alpha=1$ and $\alpha=2$ correspond to Cauchy
and Gauss distributions respectively which are mostly used in the
correlation femtoscopy. For the first case the Laguerre
polynomials, $L_{n}(x)$, are used as
$\left.\{h_{n}(x)\}\right|_{n=0}^{\infty}$; the Hermite
polynomials, $H_{n}(x)$, are chosen as the closed system of
orthogonal polynomials for the second specific case
\cite{Csorgo-PLB-489-15-2000}.

The generalized parametrization (\ref{eq:5.1}) contains
the important physical information with regard of the possible high
irregular geometry of emission region and dynamics of its creation and
it is additional with respect to information derived for sets
$\mathcal{G}_{1,2}$ of space-time parameters based on traditional Gaussian
parametrization. At now there are a few studies \cite{Okorokov-PRC-71-044906-2005,*Eggers-arXiv-hep-ex-0511050,*Abelev-PLB-739-139-2014,*Abelev-PRC-89-024911-2014} which used the formalism outlined above for specific case of Gaussian distribution only. But these investigations confirm already the importance of detailed analysis of peak shape of CF.
Thus high statistics and parameters of multiparticle final state for FCC energies allow us the qualitative study of complex geometry of emission region for secondary pions.

\section{Summary} \label{sec.6}
The following conclusions can be obtained by summarizing the results of the present study.

Energy dependence is investigated for main BEC
parameters from the set $\mathcal{G}_{1}$
derived in the framework of 3D Gaussian approach in $p+p$ collisions as well as for the set
of important additional observables $\mathcal{G}_{2}$.
Analytic function is suggested for approximation of energy
dependence of main BEC parameters. The fits demonstrate statistically acceptable qualities for $\lambda$ and for most radii even for taking into account statistical errors of experimental points.
Smooth curves calculated
for energy dependence of the set $\mathcal{G}_{2}$ of additional
BEC parameters agree with corresponding experimental
data at least of qualitative level. The
estimation of emission duration of pions in $p+p$ collisions increases from about $0.4$ fm/$c$ at RHIC energy up to $1.3$ fm/$c$ at the LHC energy.
Estimations are obtained for wide set of space-time characteristics of pion source at FCC energies on the basis of the fit results for $p+p$ and nulceus-nucleus collisions. The pion source is characterized by linear sizes about 2 fm in $p+p$, about 4 -- 5 fm in $p+\mbox{Pb}$ and 6 -- 9 fm in $\mbox{Pb}+\mbox{Pb}$ collisions at FCC energies. Volume of pion source at freeze-out is estimated from few hundreds of fm$^{3}$ in $p+p$ through few thousands of fm$^{3}$ in $p+\mbox{Pb}$ up to $10^{4}$ fm$^{3}$ in $\mbox{Pb}+\mbox{Pb}$ collisions at FCC energies.

The charged particle density and its critical value is investigated for high energy $p+p$ and $\mbox{A}+\mbox{A}$ collisions. The experimental dependence $n_{\mbox{\scriptsize{ch}}}^{\mbox{\scriptsize{pp}}}(\sqrt{\smash[b]{s_{\footnotesize{pp}}}})$ is almost flat and it describes by smooth curves reasonably for various parameterizations of total charged particle multiplicity. The $n_{\mbox{\scriptsize{ch}}}^{\mbox{\scriptsize{c,pp}}}(\sqrt{\smash[b]{s_{\footnotesize{pp}}}})$ decreases with energy nevertheless the estimations of critical value are larger significantly than charged particle density up to FCC energy. Therefore one can not expect the Bose\,--\,Einstein condensation and appropriate effects for secondary pions in $p+p$ collisions at FCC within the present approach. The charged particle density is noticeably larger in heavy ion collisions than that in $p+p$ at similar collision energies. The experimental dependence $n_{\mbox{\scriptsize{ch}}}^{\mbox{\scriptsize{AA}}}(\sqrt{\smash[b]{s_{\footnotesize{NN}}}})$ increases with energy in difference with $p+p$ reactions. Smooth curves calculated for various approaches of total charged particle multiplicity agree with experimental points. The situation is dramatically different in high energy nucleus-nucleus collisions with respect to the $p+p$ case. The charged particle density is in the range of estimations of critical parameter within its large uncertainties from the RHIC energy $\sqrt{\smash[b]{s_{\footnotesize{NN}}}}=0.2$ TeV up to the LHC $\sqrt{\smash[b]{s_{\footnotesize{NN}}}}=2.76$ TeV which is highest for experimentally available BEC results in nucleus-nucleus collisions. Furthermore there is
clear enhancement of values estimated for
$n_{\mbox{\scriptsize{ch}}}^{\mbox{\scriptsize{AA}}}$ over values for critical density
$n_{\mbox{\scriptsize{ch}}}^{\mbox{\scriptsize{c,AA}}}$ at
$\sqrt{\smash[b]{s_{\footnotesize{NN}}}} \gtrsim 10$ TeV even with
taking into account large statistical uncertainty for
$n_{\mbox{\scriptsize{ch}}}^{\mbox{\scriptsize{c,AA}}}(\sqrt{\smash[b]{s_{\footnotesize{NN}}}})$.
Thus there is possibility for Bose\,--\,Einstein condensation and novel effects, in particular, pion laser in nucleus-nucleus collisions at
FCC energy. It seems the theoretical and experimental developments are essential for future progress in this field as well as for more definite conclusions due to improvement of precision for appropriate quantities.

The generalized parametrization for two-particle BEC correlation function is suggested and it takes into account the expansion in closed system
of orthogonal polynomials for general case of non-isotropic
central-symmetrical L\'{e}vy\,--\,Feldheim distribution. Possibly, the view of BEC CF can be useful for detailed study of correlation peak shape at FCC.

\begin{acknowledgments}
The author is grateful to Prof. E. K. G. Sarkisyan for fruitful
discussions and helpful comments.
\end{acknowledgments}

\vspace*{0.5cm}
\section*{Conflict of Interests}

The author declares that there is no conflict of interests
regarding the publication of this paper.

\bibliography{VAOkorokov-v3Refs}

\end{document}